\documentclass[sigconf,10pt]{acmart}

\usepackage{booktabs} \usepackage{enumitem}
\usepackage{url}
\usepackage{color,colortbl}
\usepackage{algorithm,algorithmicx,algpseudocode}
\usepackage[mathscr]{euscript}		\usepackage{graphicx}
\usepackage{epstopdf}
\usepackage{nicefrac}
\usepackage{csquotes}
\usepackage{amsmath}
\usepackage{amsthm}
\usepackage{amsfonts}
\usepackage{boxedminipage}
\usepackage{framed}
\usepackage{subfigure}
\usepackage{xspace}
\usepackage{fancyhdr}
\usepackage{multirow}
\usepackage{booktabs}
\definecolor{lightgray}{rgb}{0.95,0.95,0.95}

\setitemize[0]{leftmargin=5pt,itemindent=5pt}
\newcommand{\etal}{\textit{et~al.}}

\newcommand{\todo}[1]{}
\newcommand{\todos}[1]{}

\newcommand{\sfsize}{\fontsize{0.8\baselineskip}{0.75\baselineskip}\selectfont}

\usepackage{enumitem}
\setlist[description]{listparindent=\parindent,leftmargin=0em,itemsep=1em,topsep=0.3em,font={\normalfont\sffamily\sfsize}}
\setlist[itemize]{topsep=0.3em, itemsep=0em, leftmargin=1.75em}

\newif\iffullpaper
\fullpaperfalse			

\iffullpaper
	\newcommand{\fullpaper}[1]{#1}
	\newcommand{\shortpaper}[1]{}
\else
	\newcommand{\fullpaper}[1]{}
	\newcommand{\shortpaper}[1]{#1}
\fi

\newcommand{\psssi}{\ensuremath{\text{PS}^3\text{I}}}

\begin{document}
	\sloppy
	\title{Privacy-Preserving Randomized Controlled Trials: \\
		A Protocol for Industry Scale Deployment }

	 \settopmatter{authorsperrow=4}

	\author{Mahnush Movahedi\footnote{this does not actually show up anywhere...}}
	\affiliation{}
	\author{Benjamin M. Case}
	\affiliation{}
	\author{James Honaker}
	\affiliation{}
	\author{Andrew Knox}
	\affiliation{}
	\author{Li Li}
	\affiliation{}
	\author{Yiming Paul Li}
	\affiliation{}
	\author{Sanjay Saravanan}
	\affiliation{}
	\author{Shubho Sengupta}
	\affiliation{}
	\author{Erik Taubeneck}
\affiliation{\institution{Facebook Inc}
			\streetaddress{385 Sherman Ave}
			\city{Menlo Park}
			\state{CA}
			\postcode{94306}
		}

\renewcommand{\shortauthors}{M. Movahedi et al.}

\begin{abstract}
Randomized Controlled Trials, when feasible, give the strongest and most trustworthy empirical measures of causal effects.  They are the gold standard in many clinical, social, and behavioral fields of study.  However, the most important settings often involve the most sensitive data, therefore cause privacy concerns. In this paper, we outline a way to deploy an end-to-end privacy-preserving protocol for learning causal effects from Randomized Controlled Trials (RCTs).  We are particularly focused on the difficult and important case where one party determines which treatment an individual receives, and another party measures outcomes on individuals, and these parties do not want to leak any of their information to each other, but still want to collectively learn a true causal effect in the world.  Moreover, we show how such a protocol can be scaled to 500 million rows of data and more than a billion gates. We also offer an open source deployment of this protocol.

We accomplish this by a three-stage solution, interconnecting and blending three privacy technologies--private set intersection, multiparty computation, and differential privacy--to address core points of privacy leakage, at the join, at the point of computation, and at the release, respectively. The first stage uses the Private-ID protocol \cite{PrivateMatching} to create a private encrypted join of the users.  The second stage utilizes the encrypted join to run multiple instances of a general purpose MPC over a sharded database to aggregate statistics about each experimental group while discarding individuals who took an action before they received treatment. The third stage adds distributed and calibrated Differential Privacy (DP) noise within the final MPC computations to the released aggregate statistical estimates of causal effects and their uncertainty measures, providing formal two-sided privacy guarantees.

We also evaluate the performance of multiple open source general purpose MPC libraries for this task. We additionally demonstrate how we have used this to create a working ads effectiveness measurement product capable of measuring hundreds of millions of individuals per experiment. 

\end{abstract}

\maketitle

	\section{Introduction}
	\label{sec1}

Randomized Controlled Trials (RCT) are considered to be the most reliable form of scientific evidence since it reduces spurious causality and bias. United States Preventive Services Task Force has recognized \emph{"evidence obtained from at least one properly randomized controlled trial with good internal validity"} as the highest quality evidence~\cite{harris2001current}. 
It is common in RCT that only one party knows the \emph{opportunities}, how individuals are randomized into test and control groups. The other party only knows the \emph{outcomes} that the treatment was meant to affect (e.g., did an individual take a specific action). Each party wishes to calculate and compare aggregate statistics of how different treatment groups performed, without revealing their input data to the other party.

\begin{figure*}[t]
	\centering
	\includegraphics[scale=0.45]{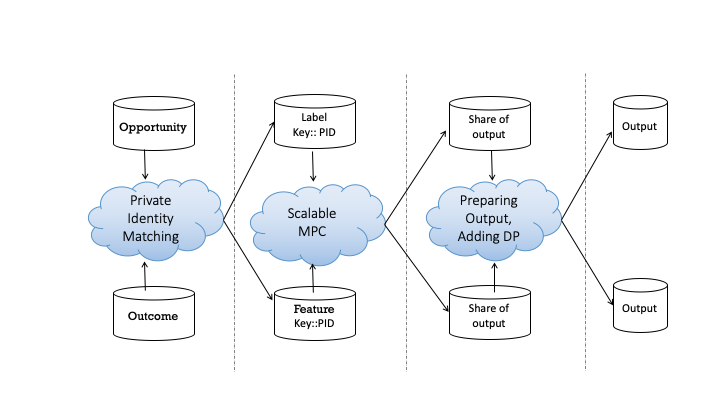} \\
	\caption{Private RCT, high level design}
	\label{fig:PLdesign}
\end{figure*}

 We design and implement Private RCT; a practical and scalable secure two-party computation system for calculating Randomized Controlled Trial experimental results. Designing a scalable privacy-preserving solution is technically challenging, especially when using one cryptographic primitive in isolation, such as either Garbled Circuit (GC) \cite{saleem2018recent} or Private Set Intersection (PSI) \cite{PrivateMatching}. Our protocol combines a suite of underlying cryptographic primitives such as GC, PSI and Secret Sharing \cite{beimel2011secret}, and incorporates them in a way that is more efficient than using any one secure computation technique by itself. Moreover, we demonstrate how to add distributed computing scaling techniques such as sharding without compromising privacy. Finally, we use Differential Privacy (DP) \cite{Dwork2013} to add noise to the output of the computation and prevent leakage of the input data based on the revealed output.
\subsection{Private Randomized Controlled Trials}
Privacy-preserving computation (secure computation) is a cryptographic method that enables parties to jointly compute a function on each of their secret inputs while preserving the privacy of the inputs. The technology guarantees that the parties will only learn the designated output of the function and they cannot access or derive each other's inputs (also known as input privacy), any intermediate values, or statistical results.

In the context of RCTs, privacy-preserving computation is highly desirable as it enables statistical measurements without giving access to the raw/un-encrypted input databases to external parties. This creates stronger guarantees to enforce the individual expectation of privacy.

Take clinical trials for example. Medical researchers randomly assign subjects into treatment and control groups and measure a few pre-determined health outcomes. While researchers can measure some demographic variables by themselves, much richer but usually unavailable demographic and behavioral data such as living conditions, mental status, social interactions, socio-economic status can be immensely valuable to improve statistical power, cut costs, and extract more credible insights from typically small-scale clinical trials. Unfortunately such auxiliary data is usually held by third parties such as hospitals, physicians, and government agencies. Private RCT solution makes it easier for these independent parties to collaborate in analyzing RCTs.

\subsection{High-level Design}

Private RCT has three main stages. The protocol is designed to make deployment in real world easy by separating the steps of the protocol and address their challenges individually. Figure \ref{fig:PLdesign} shows the overall design. 

\begin{enumerate}
	\item \emph{Private Identity Matching.}

	The first stage is to privately determine the join of users between the two parties. We use two main open source protocols for performing this Private Identity Matching: PID and PS$^3$I \cite{PrivateMatching}. The former computes a full outer join (i.e. union) of the identities using a Decisional Diffie-Hellman (DDH) based construction that is only twice as computationally expensive as standard DDH-based Private Set Intersection (PSI). PS$^3$I computes the inner-join (i.e. intersection) of the identities and generates secret shares of attached data; it uses DDH techniques and additive Homomorphic Encryption (HE). We also have ongoing work to extend both protocols to handle identity matching in cases where each row may have multiple identifiers, and we may wish to output many-to-many matches.  
	
	\item \emph{Scalable Secure Computation.} In the second stage, we calculate a pre-defined linear circuit on the output of the previous step. Since we envision different use cases, which differ in their nature of calculation, the computation step should be able to compute any circuit on the joined data. We evaluated multiple general purpose MPC frameworks for this step and concluded to build our solution based on EMP-toolkit. We used sharding mechanism to scale it from handling 2 million rows of input data to 500 million rows. We use an XOR secret sharing mechanism to hide any intermediary result in this step. 
	
	\item \emph{Differentially Private RCT Outputs.} The last step is to prepare the output. Since the output is going to be revealed to the both parties, it is important to ensure it does not break the privacy guarantees that the system promises. We need to check for access control, ensure rate limits and finally add DP noise to the output. We calibrate DP noises to ensure two-sided privacy guarantee: participant A gets only a differentially private view of B’s secret input, and vice versa. To provide differentially private confidence intervals of two-sample difference-in-means estimators, we examine several algorithms and compare their tightness, coverage, and computational cost within MPC. We also propose methods to generate DP noises distributively in the presence of semi-honest or malicious adversaries.
	
\end{enumerate}

\subsection{Our Contributions}

By combining Decisional Diffie-Hellman (DDH) PSI, Secure Multi-Party Computation (MPC), and Differential Privacy (DP), we provide a private yet performant solution to enable two parties to collaborate in RCTs. Starting from the correct underlying cryptographic primitives, we show how to use them, scale to large data sets, and handle the challenges that show themselves in practice. In addition to designing and implementing the Private RCT, we have the following contributions which can be of independent interest to the community:

\begin{enumerate}
	\item \emph{Evaluation Framework for MPC.}
	We looked into multiple services that enable secure computation to decide on the best framework that meets the requirements of Private RCT. This systematization of knowledge on MPC protocols can be of independent value for researchers and industry cryptographers. We started with a comparison across more than twenty secure computation services --- across secret sharing, garbled circuits, homomorphic encryption-based protocols. We chose ABY, EzPC, EMP-toolkit, Fancy Garbling, Scale-Mamba, Obliv-C and MPyC for a more detailed literature review --- and from them we did performance testing on ABY, EMP-toolkit, and Scale-Mamba. We will discuss the details of this evaluation in section \ref{sec:eval}.

	\item \emph{Scaling via Private Sharding.} Handling large volumes of data is important in practice. Unfortunately, most current secure computation platforms cannot handle large enough data sets at the required scale and speed imposed by our application. In this work, we designed and implemented a privacy-preserving sharding technique to make EMP scale to 500M rows. Our protocol does not reveal any intermediary information to the involved parties, i.e., the intermediary result of computation on each shard will remain private. We will discuss the details of this protocol in section \ref{sec:sharding}.

	\item \emph{Private Conversion Lift.} To test the efficiency and scalability of our design, we implemented Conversion Lift as an application that uses Private RCT in the backbone. Conversion Lift compares the actions of users in randomized test and control groups to measure the additional business driven by the advertisement. See details of implementation in section~\ref{lift}.

\end{enumerate}

	\section{Private RCT Protocol}

	In the following sub-sections, we will provide the detailed design for each step of the protocol.

	\subsection{Assumptions and Model}

	We assume there are two parties involved in the protocol Alice and Bob each have a private input $a$ and $b$ respectively. Each party has their own separate infrastructures that jointly participate in the protocol and there is no trusted third party involved. They jointly wish to compute a function $o=f(a,b)$ that is the outcome of PRC trial over their inputs and receive output $o$. 
	
	\emph{Adversarial model.} We assume both parties are semi-honest, in that they will use any information they can learn from the protocol to try and learn the other's data, but they will not deviate from the protocol specification. We assume that both parties want to learn the correct output of the computation and will not seek to poison the output by altering their inputs. Thus, the main thing we are trying to prevent is one party learning the other's inputs.  Moreover, we do consider that an adversary may craft their inputs to maximize the information they can learn passively, as discussed in \ref{sec:dp}.

\subsection{ Background on Cryptography Primitives}

\noindent \emph{\textbf{Garbled Circuit (GC)}}
Garbled circuit is a family of secure computation protocol that is more suitable for 2PC but can be extended to MPC (for small numbers). This technology is better for Boolean functions. 
In the garbled circuit, the garbler encrypts a Boolean circuit of g = f(a, ·) and sends the encrypted circuit to the evaluator. The evaluator then uses oblivious transfer (OT) to obtain the keys corresponding to the input to decrypt the garbled circuit and evaluate g(b). 
OT (the underlying technology of GC) is highly performant compared to secret sharing (on computation cost) and homomorphic encryption (on communication cost) technologies.

\noindent \emph{\textbf{Secret Sharing (SS)}}
In secret sharing, one party (called the dealer) distributes a secret among a group of parties, each of whom is allocated a share of the secret. Each share reveals nothing about the secret to the party possessing it, and the secret can only be reconstructed when a sufficient number of shares are combined together.
In secret sharing based MPC, each party creates secret shares of its input data and shares that among all parties. Then, all parties perform some intermediate computation on the received shares to get a share of the output. In the last stage, all parties will reconstruct the final output together by sending the intermediate outputs to each other.

\noindent \emph{\textbf{Homomorphic Encryption (HE)}}
A fully homomorphic encryption (FHE) scheme performs secure computation over encrypted data without decrypting it. 
Current techniques are still slow and can only evaluate small circuits. This restriction is primarily due to noise management techniques (such as bootstrapping) used to deal with a noise term in ciphertexts that increases slightly with homomorphic addition and exponentially with homomorphic multiplication.
Thus, this technique is suitable for special case circuits with more additions and fewer consecutive multiplications.

	\subsection{Private Identity Matching}
Before computation can be done for Private RTCs the two data sets must be joined by some identities of the users.  There are three matching cases we consider:
 \begin{enumerate}
 	\item  \textbf{Single unique identifier.} The simplest situation is that each user has one identifier (e.g. email) which is unique to them in one of the party's input sets. We want to compute on users who have the same identifier in both party's input sets.
 	\item  \textbf{External identifier.} In this situation, there is a common identifier shared between the two parties and each party has a many-to-many relationship between their own user ids and the external ids.
 	\item  \textbf{Many identifiers.}  Each user may have multiple identifiers (e.g. email and phone).  This case also results in a many-to-many matching between users from both parties.
 \end{enumerate}

The paper \cite{PrivateMatching} introduces two protocols for handling the case of a single unique identifier. Both protocols can compose with general-purpose MPC to enable arbitrary computation on the joined data.
The first variant which we call Private-ID (PID), allows the parties to privately compute a set of pseudorandom universal identifiers (UID) corresponding to the records in the union of their sets, where each party additionally learns which UIDs correspond to which items in its set but not if they belong to the intersection or not. This new formulation enables the parties to independently sort their UIDs and the associated records and feed them to any general-purpose MPC that ignores the non-matching records and computes on the matching ones.  The Private-ID protocol has the advantage that it only needs the identifiers from the records as input to produce the UIDs and hence for each application, parties can assemble a possibly new set of features/labels per identifier for the downstream computation without re-executing the protocol. 

 The second protocol from \cite{PrivateMatching} called Private Secret Shared Set Intersection ($\psssi$), is a natural extension of PSI where instead of learning the plaintext matched records, parties only learn additive shares of those records which they can feed to any general-purpose MPC to execute the desired computation on. The construction is based on efficiently extending existing DDH-based PSI using any additive homomorphic encryption scheme.
 The advantage of $\psssi$ over Private-ID is that its output size and hence the complexity of the subsequent MPC is proportional to the size of the intersection which in some cases is much smaller than size of the union of the two original datasets. Its disadvantage, similar to prior work, is that full records and not just the identifiers need to be ready at the time of execution, and it requires a rerun when associated records change for the same identifiers.
We have further generalized these protocols to work in matching scenarios (2) and (3) where the mapping is many-to-many. There are two ways to resolve these many-to-many mappings:
  \begin{enumerate}
  	 \item \textbf{Resolve to a single match.} In the matching case where each user has many identifiers, we can define rules of priorities or weights so as to choose the best possible match.  Both Private-ID and $\psssi$ can be generalized to do this. In cases where the many-to-many mapping results from a shared external identifier, it is generally the case that each occurrence of the identifier is of equal value in defining a match, so the next collecting approach may be better.
  	 \item \textbf{Collect many-to-many matches.}  In some applications of Private RCT it is actually best not to resolve to a single match but rather to collect all the matches into a many-to-many matching or many-to-one matching. The $\psssi$ protocol generalizes more naturally to do this than Private-ID, and we call this variant Collecting PS3I. 
  \end{enumerate}

One implication of a many-to-many identity matching is that it affects the RCT validity. If one party partitions users into Test and Control groups, it is possible that one user from the other party's set will match with users from both the Test and Control groups. We call such users contaminated and drop them from the study when running Collecting PS$^3$I on both the test and control groups simultaneously and looking for overlap in the encrypted intersections.  When running a many-identifier version of PID, we cannot drop these contaminated users, but we can count how many of them there are and use this to inform our confidence in the results.

\subsection{Private RCT Computation} \label{sec:sharding}

\begin{figure*}[h!]
	\centering
	\includegraphics[scale=0.4]{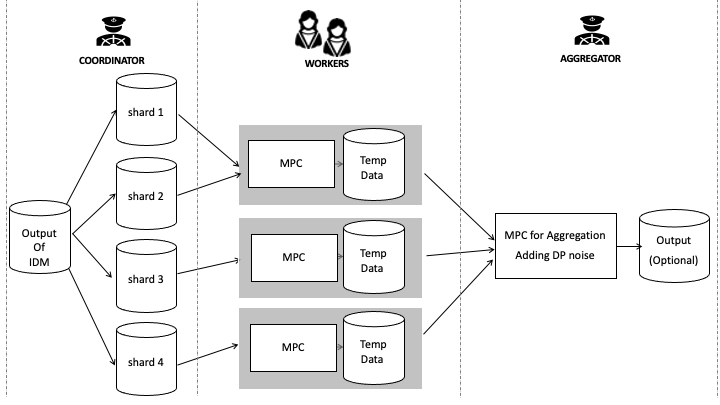} \\
	\caption{ Private RCT design. Garbler and Evaluator have the same overall architecture. }
	\label{fig:design}
\end{figure*}

We estimate that a large RCT, as is seen in advertising conversion lift studies, may routinely have up to 500M rows; thus handling large volumes of data is important in practice.
Unfortunately, EMP-toolkit and similar MPC platforms were not able to handle such a large volume of data at once in the required scale and speed imposed by the application.
EMP-toolkit was limited to 2M rows of integer data when we tested on the largest available AWS Fargate option (30GB), although it was able to handle 4M rows on a larger EC2 instance (64GB).
 
In this section we go over our design for computing a RCT game privately  across multiple containers via our privacy-preserving sharding mechanism.
In our protocol, data from the private id match is deterministically partitioned into separate container shards.
Each shard then performs a 2PC with the corresponding shard controlled by the other party.
The intermediary output of each shard remains private, i.e. the garbled values will not be open to the parties.
Instead, each party will learn an XOR share of the value.
Consequently, the aggregation step will happen after the data is reconstructed from the XOR-shares in a garbled circuit and only the final result of the computation will be revealed.

Figure \ref{fig:design} shows the high-level design of the Protocol, as it is replicated on both the Garbler and Evaluator's sides. 
Both Garbler and Evaluator each have one Coordinator, multiple Workers and one Aggregator.
The Coordinator partitions the input database into shards in a round robin method, and assigns each shard to a Worker.
Workers will evaluate the sub-circuit on the input shard and send the intermediary result to the Aggregator who evaluates the final output and adds DP noise to it. For each Worker on the Garbler side, there is a corresponding Worker on the Evaluator side that has point-to-point connection with each other and work together to evaluate the garbled circuit. Similarly, the Aggregator on the Garbler side has network connection and computes the aggregation jointly with the Aggregator on the evaluator side.

\subsubsection{Private Partitioning and Aggregation.}
To ensure no party learns the output of computation on one shard, we do not reveal the intermediary output at the end of their game.
Instead, The server will choose a random number as a new input and XOR the result of the computation with that random number in the garbled circuit.
At the end of the game, Evaluator knows the result of computation XORed with that random number and Garbler knows the random number.
Jointly, they can reconstruct the result but individually, they do not have any information about it due to the security of one-time pads.
 	\subsection{Differential Privacy in MPC} \label{sec:dp}

Deterministic MPC output can leak input information at the individual level. Typical privacy attacks on aggregated statistics include re-identification attacks, database reconstruction, and membership inference \cite{Dwork2017}. To further enhance privacy of the protocol, specifically, to make sure neither party learns any information that they don’t already know about any individual, we use Differential Privacy (DP) to add randomness to the output. See \cite{Dwork2013} for common DP definitions and algorithms. In this work, we use a more recent variant of DP definition called zero-concentrated DP (zCDP) from \cite{Bun2016}.

We aim to protect against attacks to reveal an individual’s membership, treatment status, or any outcome to any party that does not already have the data. Treatment status can itself be highly sensitive (e.g., in a clinical trial), but we also want to emphasize that membership/eligibility in the experiment and the outcome value are equally the focus of protection guaranteed by the DP mechanism.

We address three specific considerations to implement DP for Private RCT below.

\subsubsection{DP Confidence Intervals (CIs)}

Valid statistical inference with RCTs requires valid measures of uncertainty (e.g. Confidence Intervals). Valid CIs consider both the sampling variation and the randomness added by the DP noise. We consider three criteria for CIs. 1) CIs should be differentially private. 2) CIs should have the correct statistical coverage. For example, 95\% CIs should cover the true popular value at least 95\% of the time. 3) CIs should be narrow, given the DP guarantee and the correct coverage.

There are several additional challenges to construct DP confidence intervals in Private RCTs. 1) Most proposed methods for DP confidence intervals aim at estimating unknown parameters from known parametric distributions, e.g. releasing the DP mean estimate and its DP standard error from known distributions \cite{karwa2017finite, Awan2019, du2020differentially, evans2019statistically, ferrando2020general}. In RCTs we estimate the difference-in-means of two samples from unknown distributions, a context considered in \cite{dorazio2015differential}. 2) Nonparametric methods for confidence intervals are typically based on resampling techniques such as bootstrapping, which can be computationally expensive in MPC \cite{brawner2018bootstrap, covington2021unbiased}.

We compared several parametric and non-parametric approaches to construct DP confidence intervals for RCT, and found that a parametric approach based on approximating the distribution of the estimator's sampling distribution strikes a good balance for our use case with large samples. See Algorithm \ref{algo:ad} for detailed description of the algorithm.

\begin{algorithm}
  \caption{Differentially Private RCT}
  \label{algo:ad}
\begin{flushleft}
  \textbf{Input:}
  \begin{itemize}
\item $x_T$: user-level outcomes for the test group
    \item $x_C$: user-level outcomes for the control group
    \item $R$: upper bound of user-level outcomes (lower bound = 0)
    \item $\rho_1$: zCDP privacy budget for point estimate
    \item $\rho_2$: zCDP privacy budget for standard error
    \item $\alpha$: significance level of confidence interval (e.g., 10\%)
  \end{itemize}
  \textbf{Output: } $\left[\text{DP lift} - w, \text{DP lift} + w\right]$ confidence interval
\end{flushleft}

  \begin{algorithmic}[1]
    \State Clamp/Winsorize: \[
      Y_{i}=\begin{cases}
        X_{i} & \text{if }X_{i}\le R \\
        R     & \text{if }X_{i}>R.
      \end{cases}
    \]
    \State Calculate sample means, variances, and counts: $\bar{y}_T$, $\bar{y}_C$, $s^2_T$, $s^2_C$, $n_T$, $n_C$.
    \State $\text{lift} \gets \bar{y}_T - \bar{y}_C$.
    \State Standard error of lift: $se_{\text{lift}} \gets \sqrt{s_T^2 / n_T + s_C^2 / n_C}$.
    \State Sensitivity of lift: $\Delta_\text{lift} \gets \frac{R}{n_T} + \frac{R}{n_C}$.
    \State Sensitivity of the standard error of lift: $\Delta_{se_{\text{lift}}}\gets \sqrt{\frac{N^* - 1}{N^{*3}}}R$, where $N^* = \min \left(n_T, n_C\right)$.
    \State Draw scalar random noise $Z_1 \sim \text{Normal}\left(0, \frac{\Delta^2_{\text{lift}}}{2\rho_1}\right)$, $Z_2 \sim \text{Normal}\left(0, \frac{\Delta^2_{se_\text{lift}}}{2\rho_2}\right)$.
    \State DP lift $\gets \text{lift} + Z_1$, where $Z_1 \sim \text{Normal}\left(0, \frac{\Delta^2_{\text{lift}}}{2\rho_1}\right)$.
    \State DP $se_\text{lift} \gets se_\text{lift} + Z_2$, where $Z_2 \sim \text{Normal}\left(0, \frac{\Delta^2_{se_\text{lift}}}{2\rho_2}\right)$.
\State $w = \sqrt{\left(se_\text{lift} + Z_2\right)^2 + \frac{\Delta^2_{\text{lift}}}{2\rho_1}} \cdot z_{1-\alpha/2}$, where $z_{1-\alpha/2}$ is the critical value of standard normal at $1-\alpha/2$.
\end{algorithmic}
\end{algorithm}

\subsubsection{Two-side Privacy Guarantees}

In Private RCTs, each party’s input data is masked from the other party, thus DP noises need to be calibrated to ensure two-side privacy guarantee: Party A gets only a differentially private view of Party B's secret input, and vice versa \cite{Mcgregor2011}. Recall that different parties hold different parts of the data, for example, party A has the treatment variable and party B holds the outcome variable. The sensitivity of the calculation with respect to the treatment variable is larger than the sensitivity with respect to the outcome variable. It suggests that to protect party A's input data, more noise needs to be added to the final output exposed to party B. In Algorithm \ref{algo:ad}, we simplify the exposition by taking the larger sensitivity of the two parties, thereby showing both party the same results, but in practice parties may see different results due to different scales of the DP noise.

\subsubsection{Distributed DP Noise Generation} \label{sec:babble}

As a first straw man, we consider a simpler semi-honest version of the problem in which only Bob will see the output. 
Alice can generate random noise $e_a$ from any desired distribution based on the DP protocol 
and adds it as an extra input to MPC which calculates $f(a,b) + e_a$ and sends the sum to Bob where  $f(a,b)$ is the intended output of the protocol . 
However, a malicious adversary may intentionally choose a huge noise to destroy the other party’s data utility. To address this concern, we propose using the cut and choose method. Protocol asks Alice to generate a vector of $k$ random noises
$\vec{e_a}$ sampled from the desired distribution based on DP protocol and inputs it as an extra input to MPC. Bob will choose an index $i$ and parties jointly calculates $f(a,b)+\vec{e_a}[i]$ in MPC and reveal the sum to Bob. MPC also reveals all the elements in $\vec{e_a}$ other than the $i$th element which we denote as $\vec{e_a[-i]}$ to Bob. Note he can check if ${\vec{e_a}}[- i]$ has correct distribution based on DP-parameters and abort if it is not correct. Thus, our protocol has covert security. The same procedure, with reversed roles, can be done to create Alice’s output. MPC creates Bob’s and Alices’s outputs separately and reveal it only to the corresponding party.

	\section{Private Conversion Lift: Implementing a Real World Example}
	In an open source deployment of the protocol \cite{facebook2021FBPCFFacebook}, we implemented Private Conversion Lift which is used to measure the incremental (causal) impact of an advertising campaign on consumer conversions, actions such as purchases and website registrations. The party serving ads (publisher) randomly splits eligible users into test or control groups. When there is an opportunity to serve an ad from the campaign, the ad is displayed if that user is in the test group, but not displayed if they are in the control group. In both cases, the ad server logs that the user had the opportunity to see the ad. At the completion of the study, the protocol combines both the opportunity data that the publisher has recorded with outcome data from the advertiser. From these two datasets, the number of conversions in both test and control groups is counted, and the statistical estimates are made to understand incremental outcomes from the ad campaign. More details about Conversion Lift can be found in \cite{conversionlift}.

\vspace{10pt}  \noindent \textbf{Identity Matching Step.}
Private Conversion Lift is a two party protocol between an advertiser and a publisher with the following input data. The identity matching we describe here is for a single identifier $x$ (e.g. email) which is unique in each party's input data and uses the PID protocol.

The Publisher and Advertiser have an identifier $x$ for each of their users which they input into the PID protocol.  The output of the PID protocol revealed to both parties is a set of pseudorandom identifiers the size of the deduplicated union of the input sets.  Each party also receives a mapping from their users into this set of pseudorandom identifiers (which we call the "identity spine" and denote the values as UID's). But neither party learns if a particular one of their users is in the intersection or not. Both parties can then each independently sort the identity spine and align their users to it forming the full-outer join of their sets of users.

\vspace{10pt}  \noindent \textbf{Computation Step.}  We assume the size of the identity spine output from the PID identity matching is on the order of the 500 million rows that is fed to the Private Conversion Lift calculation step. For each of their users the Publisher has the common identity spine identifier UID, an opportunity timestamp, and a Boolean value to indicate if the user is in the test or control group. 
 The Advertiser also has the UID row identifier and a set of conversions for each of their users. Each conversion has associated data including a timestamp, value, and value squared that are used in the Conversion Lift calculation. 

The calculation is implemented in two games; the Lift game runs within Worker calculates the Lift results on the shard in garbled circuit and outputs the intermediary output in XOR format. The Lift game first compares the conversion's timestamp and opportunity timestamp and adds the conversion value to the total output value if the conversion happened after the opportunity. The Aggregation game runs within Aggregator that receives the intermediary outputs, XORs them pairwise to reconstruct the actual intermediary output in garbled circuit and then runs the aggregation function (addition) on them.

\vspace{10pt}  \noindent \textbf{Adding DP noise.}  The Aggregation game adds DP noise to the result before revealing it. Specifically, each party follows the distributed DP noise generation protocol in section \ref{sec:babble} and generates DP noises that will be added to the lift estimates and confidence intervals seen by the other party. The DP noises are scaled to the sensitivities to ensure two-sided privacy guarantees, following Algorithm \ref{algo:ad}.

 	\label{lift}
	\subsection{Evaluation Framework}

In this section, we compare the popular open source libraries systematically to choose the most suitable MPC framework for computing RCTs privately at scale. M. Hastings \etal~\cite{8835312} perform a similar evaluation between different MPC libraries for general use cases and we borrowed their docker container as the basis for running the protocols.      

We started with more than twenty MPC toolsets and a high level overview of the toolkits helped us initially narrow down to seven frameworks  to investigate further; ABY~\cite{demmler2015aby} by Demmler \etal, EzPC~\cite{chandran2017ezpc} by Chandran \etal, EMP-toolkit~\cite{emp-toolkit, wang2017faster, wang2017authenticated} by Wang \etal, Fancy Garbling~\cite{cryptoeprint:2016:969} by Ball \etal, SCALE-MAMBA~\cite{cryptoeprint:2010:514, cryptoeprint:2011:535, cryptoeprint:2011:091, cryptoeprint:2017:214}, Obliv-C~\cite{cryptoeprint:2015:1153} by Zahur and Evans, and MPyC~\cite{mypc} by Schoenmakers. We performed a detailed literature review for these seven MPC frameworks before narrowing down to three contenders. We ran performance tests using a simple RCT example and synthetic data, against the ABY, EMP-toolkit, and SCALE-MAMBA.

\begin{itemize}
	\item \textbf{Performance.} We consider the following metrics when performance testing the frameworks and the details of implementation and their full comparison are provided in section~\ref{sec:eval}. Here is a short summary.
	\begin{itemize}
		\item Data Volume: EMP-toolkit can handle up to one million rows of data per 16GB of memory (4 million rows with 64GB) in the same circuit. ABY can handle up to 100K rows of data with 16GB of memory, but breaks when the circuit size hits $2^{32}-1$ even if more memory is allocated.
		\item Time to Compute:  In single threaded tests, EMP-toolkit is about 14x faster than SCALE-MAMBA and ABY is about 45x faster than SCALE-MAMBA.
		\item Lag Tolerance: Sub-millisecond latency does not dramatically improve performance, but even moderate latency (60ms) can result in a 7-10x slowdown compared to 1ms.
		\item Parallelization Speedup: When we extend EMP-toolkit with our sharding design, the new implementation is about 81x faster than single threaded SCALE-MAMBA.
	\end{itemize}

	\item \textbf{Optimization.} EMP-toolkit identifies and fixes bottlenecks in various building blocks for secure computation that makes it more suitable compared to schemes without any optimization such as Obliv-C.
	As an example, EMP-toolkit uses Streaming SIMD Extensions (SSE) to improve the performance of oblivious-transfer extension, and improves the efficiency of the XOR-tree technique to avoid high (non-cryptographic) complexity when applied to large inputs. Theoretically, their optimizations reduce the cost of processing the circuit evaluator’s input by 1000x for 216-bit inputs, and even more for larger inputs. ABY use transition between  three types of sharing --- Arithmetic, Boolean, and Yao --- to improve performance. Our Private RCT cannot handle multiple types of sharing and we see this as a potential future direction.

	\item \textbf{Ease of Use.} SCALE-MAMBA is remarkably easy to use since the MAMBA language is similar to python. It also benefits from rich documentation and an active mailing list. EMP-toolkit has ample code samples and is an easy language, but does not have a well documented source code.
	It was very difficult to implement the lift game in the Fancy-Garbling framework which is a potential blocker for future developments as well. To use ABY, the protocol should be described in the circuit model for computation, which can be difficult.

	\item \textbf{Architecture and Extensibility.} For each framework, we look into its architecture to evaluate if it is modular enough to extend in the future.
	The EMP-toolkit has a large protocol set that is highly modular since each protocol is usable independently.
	The backend cryptography library of EzPC is ABY and as a result ABY is more modular compared to EzPC and its model is easier to follow and extend. EzPC extends ABY with machine learning functionality that is not immediately useful to RCTs.

	\item \textbf{Multiparty Support.} It is most common to evaluate RCTs between two parties, but there are realistic scenarios for 3 or more parties contributing data. EMP-toolkit and SCALE-MAMBA work in both two party and multiparty settings. ABY works only in the two party setting. Frigate~\cite{mood2016frigate} , PICCO~\cite{zhang2013picco} and TF-encrypted~\cite{TFEncrypted} cannot handle the two party case, thus we did not look into them deeply.

	\item \textbf{Special Dependencies.}
	EMP-toolkit and ABY do not require any special hardware. In comparison,
	SCALE-MAMBA has a HSM hardware-dependent setup phase for key generation. EzPC requires trusted execution environment such as SGX to be secure against malicious adversaries.

	\item \textbf{Completeness.}
EMP-toolkit provides a complete solution to secure computation problem including an efficient compiler and primitive cryptography protocols (circuit garbler and OT). Obliv-C's focus is on improving the efficiency of compiler, but it runs on a very simple and not optimized underlying circuit garbler and OT.
	Frigate~\cite{mood2016frigate}, CBMC-GC~\cite{10.1145/2382196.2382278} and FRESCO~\cite{FRESCO} are the only compilers and do not provide a complete secure computation service.
MPyC is suitable for rapid prototyping and teaching but not for production.

\end{itemize}
 	\section{Performance Evaluation Results} \label{sec:eval}
In this section, we present performance characteristics of a simple lift game in SCALE-MAMBA, ABY and EMP-toolkit, including tests for high data volume and high latency between nodes. We also show the speedup on EMP-toolkit from running multiple processes per container, and multiple containers. We ran all performance tests on AWS EC2 instances with 64 GB of memory and 16 CPUs unless otherwise noted.

\subsection{Performance Testing}
We began by testing the lift game with each party only using a single container. We ran each experiment ten times and provide the average among all of them as the result.

\subsubsection{Performance against input size}
\begin{figure}[h]
	\centering
	\includegraphics[scale=0.5]{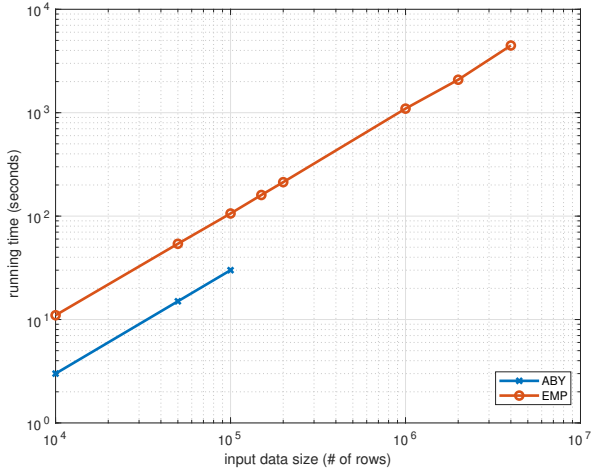}
	\caption{ Wall time for ABY and EMP-toolkit at different data volumes.}
	\label{fig:volume}
\end{figure}

We tested ABY and EMP-toolkit for different data volumes with results shown in Figure \ref{fig:volume}.  We discovered that both scaled linearly in time with increased data volume. We were able to scale EMP-toolkit by adding more memory (about 16GB per 1M rows), but we were unable to scale ABY with more memory. ABY failed with an out of memory error between 100K and 110K rows of data, regardless of how much memory was allocated. We ran input size tests in the same AWS Availability Zone.

\subsubsection{Performance against network delay}
\begin{figure}[h]
	\centering
	\includegraphics[scale=0.5]{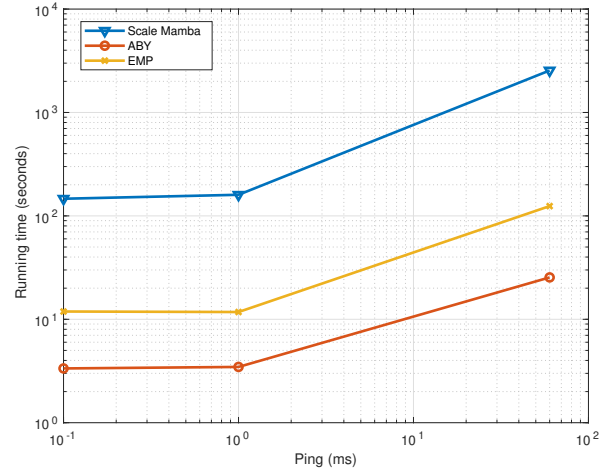}
	\caption{ Wall time for ABY, EMP-toolkit, and SCALE-MAMBA for different ping times.}
	\label{fig:ping}
\end{figure}

We then tested the same game but located the different parties in different AWS zones/regions; results shown in Figure \ref{fig:ping}. For lowest latency, we placed the parties in the same Availability Zone (0.1 ms average latency). We then placed them in different Availability Zones in the same region (1 ms average latency). We also tested different AWS regions (60 ms average latency). We found that all three protocols experienced severe performance degradation with even modest network latency. We ran network delay tests using 10K input rows.

\subsection{Performance of the Protocol at Scale}
We ran additional performance tests to measure scalability of conversion lift game in EMP-toolkit. 

\subsubsection{Scaling Up --- Local Concurrency}
We tested the effect of running simultaneous games from the same hardware. We used 4M rows of data total in each case --- more simultaneous games means each game has to calculate fewer rows, e.g., with 8 games each game only needs to calculate 500K rows. The running time of each case is shown in Table \ref{scaleuptable}. The number represents amortized time which is calculated as sum of all running time across instances divided by the number of game instances ran.
\begin{table}[h!]
	\centering
	\footnotesize
	\begin{tabular}{| c | c | c | c | c | c |}
		\hline
		\# of concurrent calculations &  1 & 2  & 4 & 8 & 16  \\
		\hline
		running time (min)  & 74.43 & 40.06 & 21.79 & 14.94 & 12.77   \\
		\hline
	\end{tabular}
\vspace{5pt}
	\caption{Running time against different numbers of concurrent game instances with fixed total input size.}
	\label{scaleuptable}
\end{table}

Table \ref{scaleuptable} shows that the running time decreases sub-linearly with the number of concurrent running games, with the most efficient gains coming from a concurrency of 4 games.

\subsubsection{Scaling Out --- Distributed Calculation}
We tested the effect of using many containers simultaneously via our new private sharing mechanism to process a very large RCT with 500M rows of data. We used AWS Fargate containers with 4 CPU and 30GB memory. Each container ran single-threaded. We ran the experiment three times and the average running times with 10 and 50 containers are listed in Table \ref{scaleouttable}.  The results show that the running time decreases linearly with the number of simultaneous containers per party, demonstrating negligible coordination overhead.
\begin{table}[h!]
\centering
\footnotesize
\begin{tabular}{| c | c | c |}
\hline
\# of containers on each party &  10 & 50   \\
\hline
running time (min)  & 1524 & 252   \\
\hline
\end{tabular}
\vspace{5pt}
\caption{Running time against different numbers of Fargate containers on each party.}
\label{scaleouttable}
\end{table}

\subsubsection{Private Conversion Lift, Results}
We ran a conversion lift study with 500M rows of synthetic data which was sharded into 50 shards of 10M rows. 
We let each worker handle one shard (10M rows) and we leveraged one coordinator to handle 50 workers.
In the test run, the publisher called the coordinator that started 50 Fargate containers in an Elastic Container Service (ECS)  in a Virtual Private Cloud (VPC) that is connected to the advertiser's VPC through VPC peering. Once the publisher's workers are running, it outputs the IPs of these workers which we shared with the advertiser. The advertiser then called its coordinator to start 50 Fargate containers that run conversion lift along with the publisher's containers. Total running time was 4hr 12min, and the AWS bill of the study is estimated to be \$1600.

\section{Related Work}
\label{related}

The closest line of work to this paper are protocols that compute a function on the intersection of two private sets also known as join and compute protocols. Circuit-based constructions such as~\cite{huang2012private, pinkas2015phasing, pinkas2018efficient, pinkas2019efficient} support arbitrary computation on the intersection by reducing the problem to executing private equality tests using a general-purpose MPC protocol. These constructions are complex, have larger communication costs, but are more generalizable.

Another method to solve the problem is to extend classic PSI wherein two parties compute and reveal the intersection without revealing anything else about the two sets. Custom DDH-style protocols focus on computing the cardinality or linear functions of the intersection~\cite{de2012fast, ion2020deploying}. They are simpler and more communication-efficient but so far have only enabled a limited set of computations on the intersection.

Our design combines the DDH-style and the circuit-based approaches, in that we take advantage of the fast communication-efficient DDH protocol which can scale nicely when composed with a lightweight yet general-purpose downstream circuit based computation.

There is a small but important literature on differentially private statistical inference, which we advance on \cite{dorazio2015differential, karwa2017finite, brawner2018bootstrap, evans2019statistically, ferrando2020general, du2020differentially, covington2021unbiased}. We have described some of this in the body of the paper.
 
	\section{Summary}
	We establish that a sharded version of  EMP-toolkit is an appropriate choice for building a system able to handle a large advertising effectiveness RCT study. Further research is necessary to develop systems capable of handling the full breadth and volume of studies performed today. MPC is a very active field of research, and we expect to see new libraries emerge, and for future improvements to some existing tool sets. We hope that this research gives context and inspires open source library developers to tackle scaling, hard memory limits, strong coupling between compilers and protocols, and general usability.

	\subsection*{Disclosure Statement}
	The authors have no conflicts of interest to declare.

	\vspace{-0.1cm}
	\bibliographystyle{ACM-Reference-Format}
	\bibliography{secbib}

	\appendix

\section{Appendix: Literature Review on MPC Frameworks}

\subsection{ABY}
\begin{itemize}
	\item \emph{Architecture.}
	The ABY framework protocol is based on three types of sharing, Arithmetic, Boolean, and Yao. In addition, they implement wrappers for conversions between Cleartexts and three types of sharings. The code structure reflects this architecture and is easy to follow. The ABY framework has a modular design that can easily be extended to additional secure computation schemes, computing architectures, and new operations, while also allowing special purpose optimizations on all levels of the implementation. It has good abstraction and distinction between application and framework. 
	
	\item \emph{Performance optimization.}
	Has an offline/online phase which allows for better performance/less communication in the online phase. It supports SIMD construction in c++. 
	
	\item \emph{Scalability and Parallelism.} 
	It has parallelism functionality by setting up the circuit as SIMD. They are using two threads in the setup phase in their example. 
	
	\item \emph{Developer point of view.} 
	ABY provides powerful low-level cryptographic tools to optimize the computation. The users should be familiar with underlying MPC protocols, their relative tradeoffs. The protocol should be described in the circuit model for computation.
	
\end{itemize}

\subsection{EzPC}
\begin{itemize}

	\item \emph{Architecture.} 
	The microsoft team provides EzPC and three other components for secure computation in a TensorFlow style.
	EzPc is a framework for secure 2PC.
	Athos is an end-to-end compiler from TensorFlow to a variety of semi-honest MPC protocols. Athos leverages EzPC as a low-level intermediate language.
	Porthos is a semi-honest 3 party computation protocol which is geared towards TensorFlow-like applications.
	Aramis is uses hardware with integrity guarantees to convert any semi-honest MPC protocol into an MPC protocol that provides malicious security. Notice we usually call the toolset as EzPC. However, we misuse the library name here since EzPC is just one underlying part. Each one of the above components can be used individually. The architecture is a little confusing. For example, there is a blur line between EzPC and Athos. But, not a major blocker. The backend of EzPC is ABY. They call the ABY library every time they have a private operation. They improve the way developers write code by moving away from defining circuits and support transflow language. They also have a compiler that can choose the correct underlying library between arithmetic circuit, boolean circuit Yao.
	
	\item \emph{Performance optimization.} 
	EzPC is the first to generate protocols that combine both arithmetic sharing and garbled circuits for better performance. It can understand the data type (integer or boolean) of the operation and call the appropriate library to create arithmetic or boolean circuit respectively. 
	
	\item \emph{Scalability and Parallelism.} 
	EzPC addresses the scalability concern using “secure code partitioning” technique: it decomposes the original program into a sequence of smaller programs.
	
	\item \emph{Developer point of view.} 
	The protocol should be defined in transflow which can be more limited than Scale-Mamba but easier than the circuit model of ABY.
	
\end{itemize}
\subsection{EMP-Toolkit}
\begin{itemize}
	\item \emph{Architecture.}
	EMP-toolkit has the following modules:
	\begin{itemize}
		\item Oblivious-transfer protocol family
		\item Oblivious-transfer protocol family
		\item Input/Output for different types
		\item Basic operations
	\end{itemize}
	
	The toolkit lacks a well-defined high-level architecture. It is a set of mostly independent implementations of different 2pc and MPC protocols. emp-ag2pc is an interesting one which implements a constant-round protocol for 2pc boolean circuits, that is secure against malicious adversaries. Note, ABY is not secure in a malicious setting and EzMPC requires SGX.
	
	\item \emph{Performance optimization.} 
	Different protocols in the suit have different performance levels.
	
	\item \emph{Scalability and Parallelism.} 
	As an experiment developers of EMP-toolkit executed their protocol with a large number of parties located all over the world, computing AES with 128 parties across 5 continents in under 3 minutes.
	
	\item \emph{Developer point of view.} 
	EMP-toolkit has an effective circuit generation tool and supporting libraries (OT) which makes it easy to work with for circuit generation. However, the end-to-end flow is not seamless. 
\end{itemize}
\subsection{Fancy Garbling/Swanky}
\begin{itemize}
	
	\item \emph{Architecture.}
	Swanky is a suit of Rust libraries for MPC that contains the following protocols:  
	
	The positive aspect of this suit is that it has internal PSI protocols.
	
	\item \emph{Performance optimization.}
	For arithmetic circuits over the integers, the construction results in garbled circuits with free addition, weighted threshold gates with cost independent of fan-in, and exponentiation by a fixed exponent with cost independent of the exponent. For boolean circuits, the construction gives an exponential improvement over the state of the art for threshold gates (including AND/OR gates) of high fan-in.
	
	\item \emph{Scalability and Parallelism.}
	They focus improvement on threshold computations with very high fan-in.
	
	\item \emph{Developer point of view.}
	The library codebase looks professional but we did not implement an example yet.
\end{itemize}

\subsection{Scale-Mamba}
\begin{itemize}
	
	\item \emph{Architecture.}
	It has two main protocols: 
	SCALE is the executive part of the framework and itself has two sub-systems which are fully integrated
	offline phase, 
	online phase,
	MAMBA Compiler.
	The compiler takes a program written in our special language MAMBA, and then turns it into byte-code which can be executed by the SCALE system.
	A potential red flag is that the setup phase for key generation is not secure by its own and requires using HSMs to generate keys. However, we should investigate this for other frameworks in general.
	
	\item \emph{Performance optimization.}
	The offline phase will do a heavy computation to make the online phase faster. Thus, using the system to run very small programs in the online phase is going to be inefficient. Optimizations efforts involve: 
	multiplication triples,
	square pairs, 
	shared bits.
	
	\item \emph{Scalability and Parallelism.}
	Using multiple threads enables users to get high throughput. The MAMBA compiler decides on the number of online threads based on the program. There are also helper threads for the offline phase. 
	
	\item \emph{Developer point of view.}
	It has an average quality code base with minimal support. 
\end{itemize}

\subsection{Obliv-C}

\begin{itemize}
	
	\item \emph{Architecture.}
	The focus of this library is the language (Oblive-C) and a well suited compiler for it. Oblive-C provides a type system that incorporates oblivious data types and ensures the data remains private.
	
	\item \emph{Performance optimization.}
	The execution part is simple and not that optimized.
	
	\item \emph{Scalability and Parallelism.}
	It gets most parallelism efforts of a simple C. They implemented some threading support library to coordinate between two parties. 
	
	\item \emph{Developer point of view.}
	It is a stable library that is used in many applications. It is a little old and do not have active support.
\end{itemize}
\subsection{MPyC: Rapid-Prototyping MPC Computations in Python}
\begin{itemize}
	\item \emph{Architecture.}
	MPyC is not a production ready platform and is designed as an educational tool that enables prototyping for research. It has a good modularity.
	
	\item \emph{Performance optimization.}
	MPyC’s execution model lets the CPU(s) do useful work while waiting for data from other players.
	
	\item \emph{Scalability and Parallelism.}
	MpyC has a strong parallelism model that is its selling point. It also has execution paradigm to multiplex several “blocking” tasks to single thread.
	
	\item \emph{Developer point of view.} super easy to get it run on one machine. Haven't figured out how to run parties on multi machines. It is written in pure python which makes implementing easy. However, it was written for rapid prototyping and teaching purposes.
	
\end{itemize} 
\end{document}